\documentclass[showpacs,preprintnumbers,pre,nofootinbib,onecolumn]{revtex4}
\usepackage{amsmath}
\usepackage{graphicx}
\usepackage{dcolumn}
\usepackage{bm}

\input{tcilatex}

\begin{document}

\preprint{}
\title{Pure multiplicative stochastic resonance of anti-tumor model with
seasonal modulability}
\author{Wei-Rong Zhong}
\altaffiliation[ ]{Corresponding Author}
\email{wr-zhong@126.com}
\author{Yuan-Zhi Shao}
\author{Zhen-Hui He}
\affiliation{State Key Laboratory of Optoelectronic Materials and Technologies,\\
Department of Physics, Sun Yat-sen University, 510275 Guangzhou, People's
Republic of China}

\begin{abstract}
The effects of pure multiplicative noise on stochastic resonance in an
anti-tumor system modulated by a seasonal external field are investigated by
using theoretical analyses of the generalized potential and numerical
simulations. For optimally selected values of the multiplicative noise
intensity quasi-symmetry of two potential minima and stochastic resonance
are observed. Theoretical results and numerical simulations are in good
quantitative agreement.
\end{abstract}

\maketitle

Chemotherapy remains a traditional option for most advanced cancer.
Immunotherapy, however, is a less conventional treatment modality. Usually,
chemotherapy and immunotherapy have been regarded as unrelated, so
relatively little research has investigated the relationship between these
two therapies. Chemotherapy kills tumor cells in a special way periodically,
but immunotherapy restrains the growth of tumor cells in a more likely
linear way [1, 2]. Since these different responses of tumor cells to
chemotherapy and immunotherapy, when taken together, they imply that there
is an interesting and significative case for combining chemotherapy and
immunotherapy in tumor treatment.

More than ever, cancer research is now an interdisciplinary effort which
requires a basic knowledge of commonly used terms, facts, issues, and
concepts. In the past decade, many studies have focused on the growth law of
tumor cells via dynamics approach, specially using noise dynamics [3-9].
Phase transition of tumor growth induced by noises is one of the most novel
foundations in recent years. Another phenomenon---known as stochastic
resonance (SR) ---shows that adding noise to a system can sometimes improve
its ability to transfer information. The basic three ingredients of
stochastic resonance are a threshold, a noise source and a weak input, it is
clear that stochastic resonance is a common case and generic enough to be
observable in a large variety of nonlinear dynamical systems [10, 11]. Thus,
it is reasonable to believe that SR can also occur in a tumor dynamical
system.

The mean field approximate analysis is a conventional theory for SR. It is
originally proposed for symmetrical bistable systems with additive noise
source [12]. The improvements of the theory of SR have included monostable
systems [13, 14], asymmetrical systems [15] and double-noises systems
[16-17], but in all these studies the system has an additive noise source
and an independent external field. To pure multiplicative noise systems,
especially to many complex dynamical systems, it is still far difficult to
solve them exactly, thus numerical methods are comparatively convenient
options to solve these complex dynamical systems [18, 19].

In this letter, chemotherapy and immunotherapy are joined by an anti-tumor
model with three elements, which are (1) a fluctuation of growth rate, (2)
an immune form, and (3) a weak seasonal modulability induced by
chemotherapy. Based on the analyses on its unique stochastic differential
equation and relevant Fokker-Planck equation, we investigate a new type of
SR phenomenon of anti-tumor model through both theoretical analysis and
numerical calculation. We call this effect pure multiplicative stochastic
resonance (PMSR) to emphasize that pure multiplicative noise induces a
synchronization, which can be described by the symmetry of the potential
wells. A presupposition is given about the relationship between SR and tumor
response to treatment.

Lefever and Garay [20] studied the tumor growth under immune surveillance
against cancer using enzyme dynamics model. The model is, 
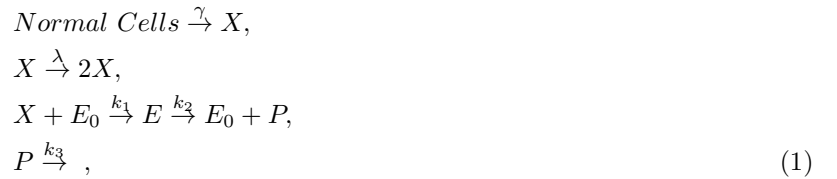
\begin{eqnarray}
&&Normal\text{ }Cells\overset{\gamma }{\rightarrow }X,  \notag \\
&&X\overset{\lambda }{\rightarrow }2X,  \notag \\
&&X+E_{0}\overset{k_{1}}{\rightarrow }E\overset{k_{2}}{\rightarrow }E_{0}+P,
\notag \\
&&P\overset{k_{3}}{\rightarrow }\text{ },  \TCItag{1}
\end{eqnarray}%
in which $X,P,E_{0}$ and $E$ are respectively cancer cells, dead cancer
cells, immune cells and the compound of cancer cells and immune cells, $%
\gamma ,\lambda ,k_{1},k_{2},k_{3}$ are velocity coefficients. This model
reveals that normal cells may transform into cancer cells, and then the
cancer cells reproduce, decline and die out ultimately. This model can be
derived to an equivalent single-variable deterministic dynamics differential
equation [5],%
\begin{equation}
\frac{dx_{n}}{d\tau }=r_{n}x_{n}(1-\frac{x_{n}}{K_{n}})-\varphi (x_{n}) 
\tag{2}
\end{equation}%
where $x_{n}$ is the population of tumor cells; $r_{n}$ is their linear per
capita birth rate and $K_{n}$ is the carrying capacity of the environment,
respectively. $\varphi (x_{n})$ quantifies the abilities of recognizability
and attack which immune cells have to tumor cells. Its form is given by $%
\varphi (x_{n})=\beta x_{n}^{2}/(\epsilon ^{2}+x_{n}^{2})$ [5, 21], here $%
\beta $ is the immune coefficient, $\epsilon $ gives a measure of the
threshold where the immune system is `switched on'. For convenience, we set $%
x=x_{n}/\epsilon ,r=r_{n}\epsilon ,K=K_{n}/\epsilon ,t=\tau /\epsilon ,$ and
obtain a non-dimension form of Eq.(2),%
\begin{equation}
\frac{dx}{dt}=rx(1-\frac{x}{K})-\frac{\beta x^{2}}{1+x^{2}}  \tag{3}
\end{equation}

Like most species, seasonal growth is a common feature of tumor cells,
especially when they are in a periodic chemotherapeutic treatment [2]. This
means that the growth rate of tumor cells should be a periodic form, for
example a cosine form. If considering the fluctuation of environment, the
growth rate $r$ in Eq.(3) should be rewritten as $r_{0}+A_{0}\cos (\omega
t)+\xi (t),$ where $\xi (t)$ is the Gaussian white noises defined as $%
\langle \xi (t)\rangle =0,\ \ \langle \xi (t)\xi (t^{\prime })\rangle
=2M\delta (t-t^{\prime })$, in which $M$ is the noise intensity. The
equivalent stochastic differential equation of Eq.(3) will be,%
\begin{equation}
\frac{dx}{dt}=r_{0}x(1-\frac{x}{K})-\frac{\beta x^{2}}{1+x^{2}}+x(1-\frac{x}{%
K})A_{0}\cos (\omega t)+x(1-\frac{x}{K})\xi (t)  \tag{4}
\end{equation}

In the absence of external field, i.e., $A_{0}=0,$ if set $%
f(x)=r_{0}x(1-x/K)-\beta x^{2}/(1+x^{2}),$ and $g(x)=x(1-x/K),$ one will
obtain the Fokker-Planck equation of Eq.(4),%
\begin{equation}
\frac{\partial P(x,t)}{\partial t}=-\frac{\partial \lbrack A(x)P(x,t)]}{%
\partial x}+\frac{\partial ^{2}[B(x)P(x,t)]}{\partial x^{2}}  \tag{5}
\end{equation}%
in which%
\begin{eqnarray}
A(x) &=&f(x)+Mg(x)g^{^{\prime }}(x)  \notag \\
B(x) &=&Mg^{2}(x)  \TCItag{6}
\end{eqnarray}

The stationary probability distribution of the system is obtained from
Eqs.(5) and (6),%
\begin{equation}
P_{st}(x)=N\exp [-\frac{U_{eff}(x)}{M}]  \tag{7}
\end{equation}%
where $N$ is a normalization constant, and the generalized potential%
\begin{eqnarray}
U_{eff}(x)=\frac{2\beta K^{3}}{(1+K^{2})^{2}}(\ln |\frac{\sqrt{1+x^{2}}}{K-x}%
| &&+K\arctan x)+(r_{0}+M)\ln |\frac{K-x}{x}|  \notag \\
&&+\frac{\beta K^{2}}{1+K^{2}}(\frac{1}{K-x}-\arctan x)+M\ln |\frac{x^{2}}{K}%
|  \TCItag{8}
\end{eqnarray}

The generalized potential $U_{eff}(x)$ versus the population of tumor cells $%
x$ is plotted in Fig.1 at different noise intensities $M$. Obviously, the
potential has two stable states, and its minima are given by $%
A(x)-B^{^{\prime }}(x)=0$, i.e., $r_{0}(1-x/K)-\beta
x/(1+x^{2})-M(1-x/K)(1-2x/K)=0$. Here the positions $x_{1}$, $x_{2}$ of the
potential minima are regarded as the inactive state and the active state of
tumor cells, respectively.The generalized potential is an asymmetric
bistable potential wells, and its potentials at $x_{1}$ and $x_{2}$ change
with the noise intensity $M$. We define the difference of the potentials at $%
x_{1}$, $x_{2}$ as $\Delta U=|U_{eff}(x_{1})-U_{eff}(x_{2})|$ and plot the
relationship between $\Delta U$ and the noise strength $M$ in Fig.2,
observing a minimum at a nonzero noise level. The figure displays that the
symmetry of the potential wells is determined by the multiplicative noise
intensity. Although they are asymmetry at high and low values of noise
intensity, the potential wells are quasi-symmetry and $\Delta U\ $tends to
zero at a suitable noise strength. This makes SR possible for the systems
with asymmetry potential wells. Due to this unique generalized potential,
the system modulated by an external field undergoes a special response to
multiplicative noise.

If inputting a seasonal signal $A_{0}\cos (\omega t)$, as shown in Eq.(4), a
time series is taken to monitor the anti-tumor system response based on a
numerical method for stochastic differential equations [19]. At low values
of $A_{0}$ and $\omega $ (i.e., $A_{0}\ll 1,\omega \ll 1$)$,$ the symmetry
of two potential wells decides the synchronization of the probability skips
between them mainly. In the presence of additive noise, the relationship
between the signal-to-noise ratio ($SNR_{1}$) and the height of the
potential barrier ($\Delta U_{k})$ is given by $SNR_{1}=4\pi Ar_{k}/D,$ $%
r_{k}=N_{k}\exp (-2\Delta U_{k}/D)$, in which $N_{k}$ is an independent
constant, and $A,$ $D$ are external field amplitude and additive noise
intensity, respectively [12]. This indicates that the height of the
potential barrier affects the existence of SR. However, in the absence of
additive noise, the above formulas of the traditional theory of SR are not
suitable for this pure multiplicative noise problem. Here the symmetry of
the potential has the main effects on the synchronized hopping between the
two states, i.e., the existence of SR is mainly determined by the symmetry
of the potential wells but not the height of the potential barrier.
Accordingly, for the sake of simplicity, we have proposed an approximate
substitute ($R_{1}$) for the signal-to-noise ratio (SNR), which has the same
form as the $SNR_{1}$ mentioned above and is simply written as%
\begin{equation}
R_{1}=N_{0}\exp (\frac{-\Delta U}{M})  \tag{9}
\end{equation}%
where $N_{0}$ is a non-dimension proportionality coefficient.

Figure 3 shows selected time series at different multiplicative noise
intensities, and their associated power spectral intensities are shown in
Fig.4. For a right value of noise intensity, the populations of tumor cells
skip back and forth between active state and inactive state, indicating the
synchronous response of tumor cells to the treatment at an optimal
fluctuation in their growth rate. The SNR is defined as the ratio of
intensity of the peak in the power spectral intensity to the height of noisy
background at the same frequency. After calculating the SNR, the data are
combined and presented in Fig.5. The stochastic resonance induced by
multiplicative noise is clear and marked, with the maxima of the SNR and $%
R_{1}$ approximately at $M\sim $ 0.22. This is a pure multiplicative
stochastic resonance. With an increase in multiplicative noise intensity,
the change trend of SNR matches that of $R_{1}$ very well.

In conclusion, we have investigated the stochastic resonance induced by pure
multiplicative noise in an anti-tumor system. The seasonal factor,
reflecting the influence of chemotherapy on tumor cells, is introduced into
the conventional tumor growth model under immune systems surveillance. On
the base of analyses on the asymmetry of the generalized potential, we have
given a parameter to substitute the SNR, which consists with the numerical
results very well. Our works offer a method to analyze some complex
stochastic differential equations, although they are insufficient to give an
exact description of real tumor growth. Especially, the synchronous response
of tumor cells to chemotherapy is one of the novel findings. We expect that
these analyses and numerical findings will stimulate theoretical and
experimental works to verify pure SR in real anti-tumor systems with
seasonal treatment.

This work was partially supported by the National Natural Science Foundation
(Grant No. 60471023) and the Natural Science Foundation of Guangdong
Province (Grant No. 031554), P. R. China.

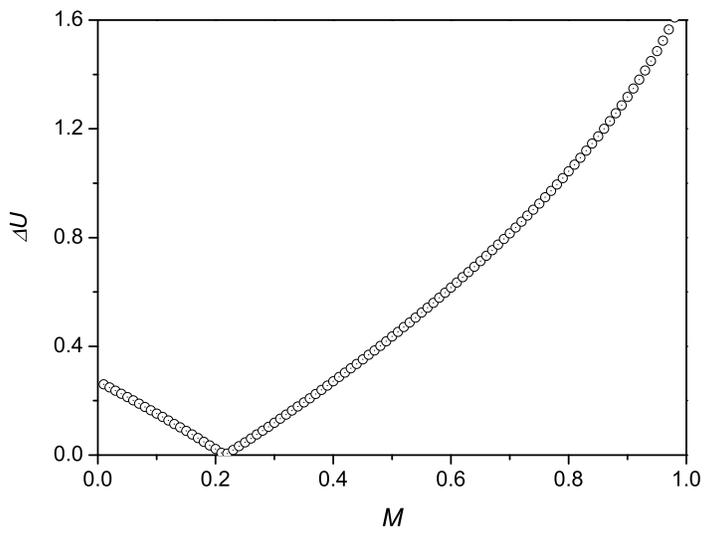
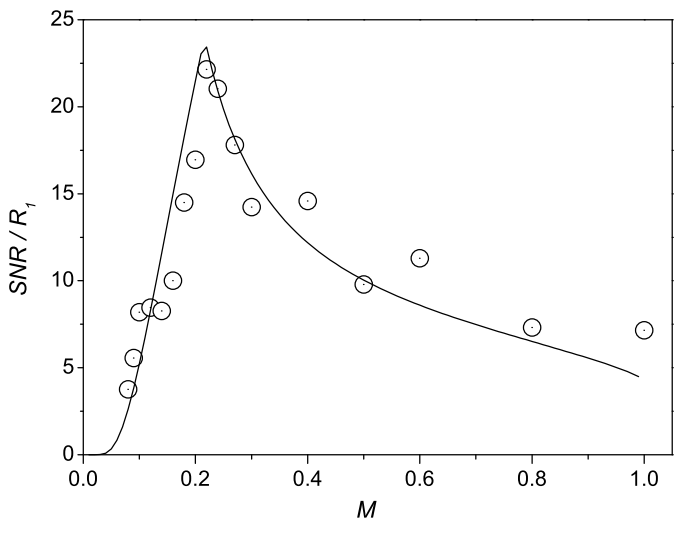
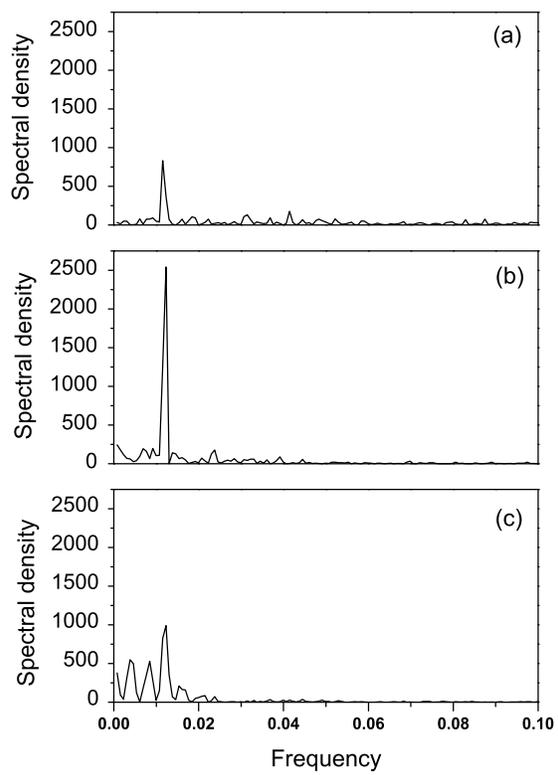
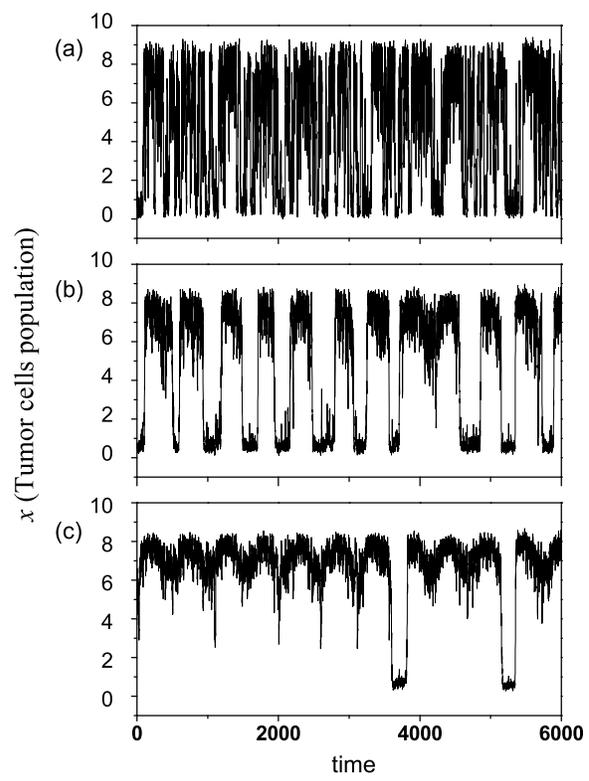
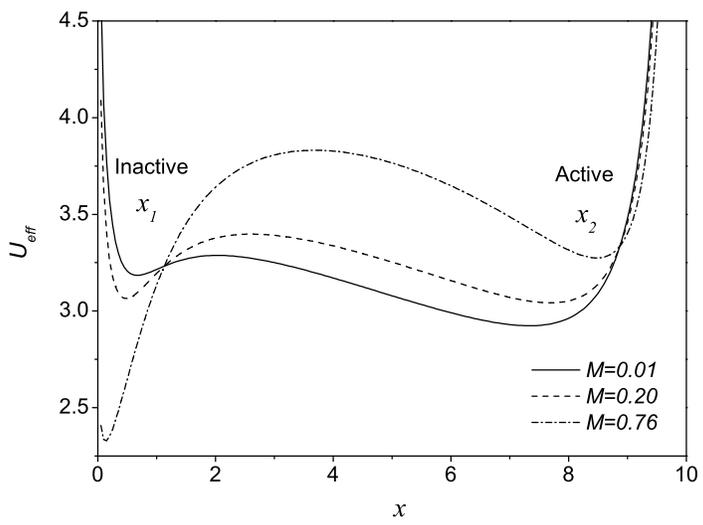
\end{document}